\documentclass[slac_one]{revtex4}
\usepackage{graphicx}
\usepackage{fancyhdr}

\begin{document}

\title{Constraining the polarized gluon PDF in pp collisions at RHIC} 

\author{F. Ellinghaus, for the PHENIX and STAR Collaborations}

\affiliation{University of Colorado, Boulder, CO 80309, USA}


\begin{abstract}

The main focus of the physics program at PHENIX and STAR that makes
use of RHIC's polarized proton beams is to figure out
how and if at all the gluons inside protons are polarized, or
to put it another way, do the
spin 1 gluons prefer to have their spins aligned or anti-aligned with
the spin of the proton, or do they just not care?  
This question is an
important part of the more general question of how the constituents of
protons, gluons and quarks, conspire to make up the overall spin 1/2
of the proton. 
Measurements of, e.g, jet and hadron, production cross-section differences
between the two cases where the two polarized protons colliding have their
spins aligned and anti-aligned 
are sensitive to the gluon polarization, which is encoded in the 
spin--dependent parton distribution function (PDF) for gluons,~$\Delta g(x)$.

\end{abstract}


\maketitle

\thispagestyle{fancy}

In the most naive model, the so-called constituent quark model, in which the protons consist of three static
quarks, two of the three
spin 1/2 quarks will align their spins with the spin of the proton,
the third one is anti-aligned, and voila, the three quarks account for
the spin 1/2 of the proton.  However, after pioneering experiments at
SLAC starting in the mid-1970's it was the European Muon Collaboration
(EMC) at CERN in the late 1980's that found the contribution of
the quark spins to the spin of the nucleon to be small and actually
consistent with zero within the uncertainties~\cite{emc}. 
The most recent results by the HERMES Collaboration at DESY~\cite{hermes_g1} and
the COMPASS Collaboration at CERN~\cite{compass_g1_d} show that the quarks
contribute about 30\% to the spin of the nucleon, leading
to our current understanding that a substantial fraction of the
proton's spin has to arise from the quark orbital angular momentum
and/or the gluons.

The basic principle of these measurements is to use deep inelastic scattering
(DIS) of longitudinally polarized leptons off
longitudinally polarized nucleons, which are usually embedded in a
gaseous or solid state target. The polarized lepton will
interact with a quark inside the polarized nucleon via the exchange of a
spin 1 virtual photon. This photon can only be absorbed by the quark
when the spins of the quark and the photon are anti-aligned due to
angular momentum conservation. The cross section for this
process can be measured for the two cases where the spin of the
leptons and the nucleons is aligned and anti-aligned, and depending on
which cross section turns out to be bigger, one can learn if the quark
spins tend to be aligned or anti-aligned with the spin of the
nucleon. 
This cross section difference is usually measured as a
function of Q$^2$, the negative four-momentum squared of the virtual
photon, and of $x$, the fraction of the momentum of the (fast) proton
carried by the struck parton, and it is given in terms of the so-called
polarized structure function g$_1(x,Q^2)$.

The possible contribution of the gluon spin $\Delta G(Q^2) = \int_{0}^{1} dx \Delta g(x,Q^2)$ 
to the nucleon spin has first been studied in DIS. The problem with studying gluons in DIS is
that the electromagnetic probe, the photon, cannot directly couple to
the gluon since gluons carry no electric charge. An indirect way of accessing the
polarized gluon PDF $\Delta g(x)$ is via
next-to-leading order (NLO) fits to the above mentioned structure function
$g_1(x,Q^2)$. 
The information on the polarized gluon PDF is contained in
the dependence of $g_1$ on $Q^2$ at a given $x$, which is nearly constant. To be
precise, it is the difference from the constant behavior, the so-called
scaling violation, which contains the information.  
However, since all
measurements of $g_1$ have been done at fixed-target experiments the
range in $Q^2$ covered is very limited, leading to rather large
uncertainties for $\Delta g(x)$. 
A more direct way to measure $\Delta g(x)$ at the DIS experiments is via a NLO
process called photon-gluon fusion (PGF), in which the virtual photon
and the gluon interact via the creation of a quark-antiquark pair. However, the
available results so far suffer from the fact that the
energy scale at the fixed-target experiments is rather low.

Thus, the motivation to study the gluon in pp collisions at the STAR and
PHENIX experiments at RHIC (Relativistic Heavy Ion Collider) at BNL is quite
natural. The involved energies are high and the 
gluon participates in leading order since the basic processes of interest are
the scattering of quarks off gluons and of gluons off gluons. 
Similar to the case of $g_1$ the cross section
difference is measured between the two cases where the two polarized protons
colliding have their spins aligned and anti-aligned. To be precise, 
the difference divided by the sum, the so-called double helicity asymmetry
\begin{equation} 
A_{LL} = \frac{\sigma^{++}-\sigma^{+-}}{\sigma^{++}+\sigma^{+-}}
= \frac{\Delta \sigma}{\sigma}, \quad \text{with} \quad
\Delta \sigma
\propto \sum_{abc}\Delta f_a\otimes \Delta f_b\otimes\Delta\hat{\sigma}^{ab\rightarrow cX'}\otimes D^{h}_c,
\label{cross_sec_asy}
\end{equation}
is measured, where the cross section $\sigma^{++}$ ($\sigma^{+-}$)
describes the reaction where both protons have the same (opposite) helicity.
The spin--dependent term is given on the rhs of Eqn.~\ref{cross_sec_asy},
where $\Delta f_a$, $\Delta f_b$ represent the spin--dependent PDFs for quarks (u,d,s) and
gluons, which, from the NLO fits to $g_1$ as mentioned above, are well known
for quarks but not well known for gluons.
The fragmentation functions
$D^h_c$ represent the probability for a certain parton $c$ to fragment into a certain
hadron $h$, and thus they are not needed in the case of jet production.
The spin--dependent hard scattering cross
sections $\Delta \hat{\sigma}$ have to be calculated using perturbative QCD.
Thus, the applicability of perturbative QCD in the kinematic regime of the $A_{LL}$
measurements is a crucial prerequisite. This has been checked by comparing
the results from NLO pQCD calculations using the well known
unpolarized PDFs for quarks and gluons with measured unpolarized cross
sections for, e.g., the inclusive production of jets~\cite{star_jets_run34} or
neutral pions~\cite{allpi0}.
Good agreement has been found giving
confidence in proceeding with measurements of the double helicity
asymmetry (Eqn.~\ref{cross_sec_asy}), which experimentally 
translates into
\begin{equation} 
A_{LL} = \frac{1}{|P_B||P_Y|}\frac{N_{++}-RN_{+-}}{N_{++}+RN_{+-}}, \quad
\text{with} \quad 
R\equiv\frac{L_{++}}{L_{+-}},
\end{equation}
where $N_{++}$ ($N_{+-}$) 
and $L_{++}$ ($L_{+-}$) 
are the experimental yield $N$ and luminosity $L$ for the case where the beams
have the same (opposite) helicity.
The relative luminosity $R$ is just the ratio of the two luminosities. 
The polarizations of the two colliding beams at RHIC are denoted by $P_B$ and $P_Y$. 
The degree of polarization is determined from the combined information of a
$\vec {p}C$ polarimeter~\cite{pC}, using an unpolarized ultra--thin
carbon ribbon target, 
and from $\vec {p}\vec{p}$ scattering, using a polarized atomic
hydrogen gas-jet target~\cite{jet}. 
The average polarization value for the data from 2005 (2006) is
$49$\% (57\%).

\begin{figure}[t]
\includegraphics[width=0.48\columnwidth]{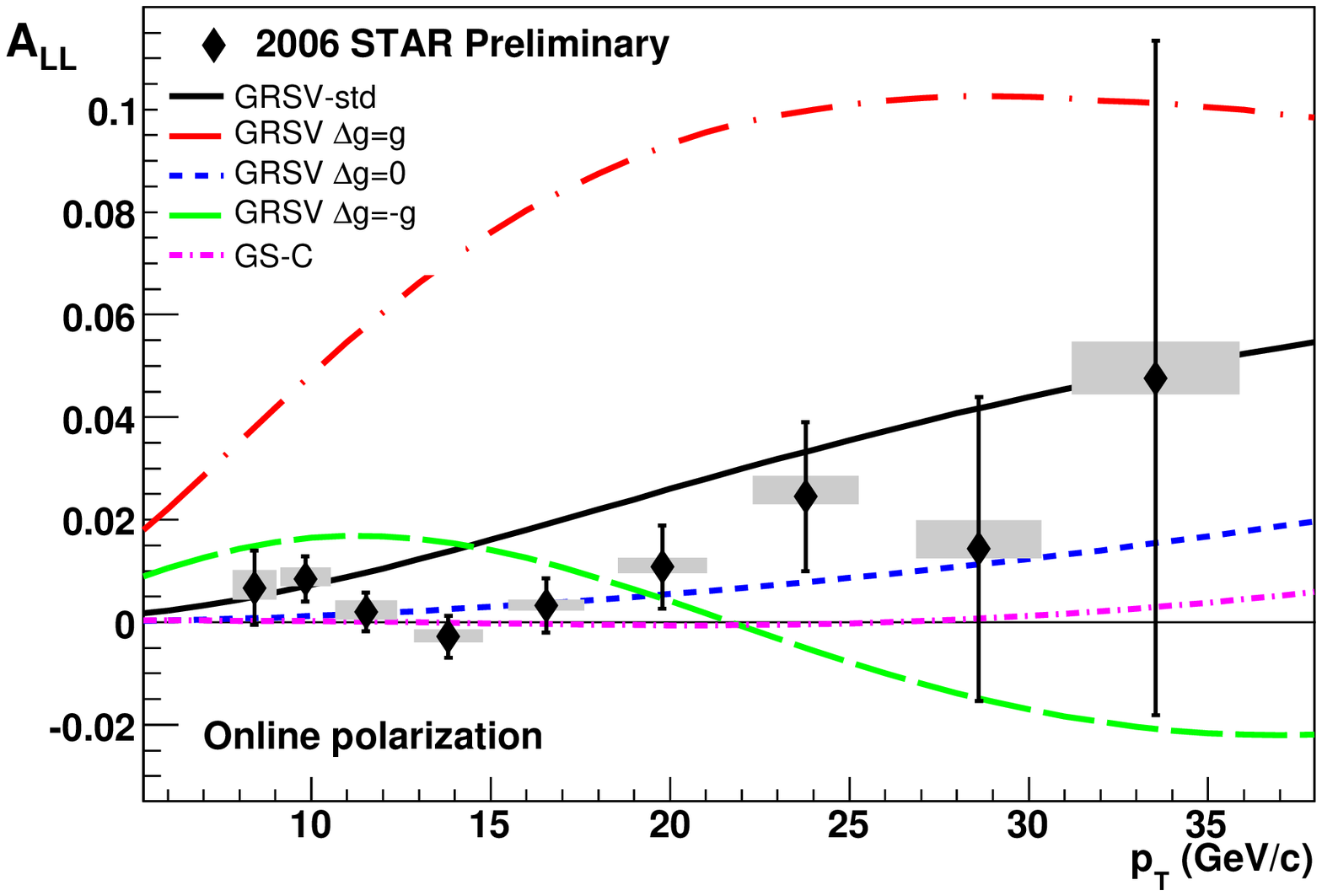}
\includegraphics[width=0.48\columnwidth, height=7.1cm]{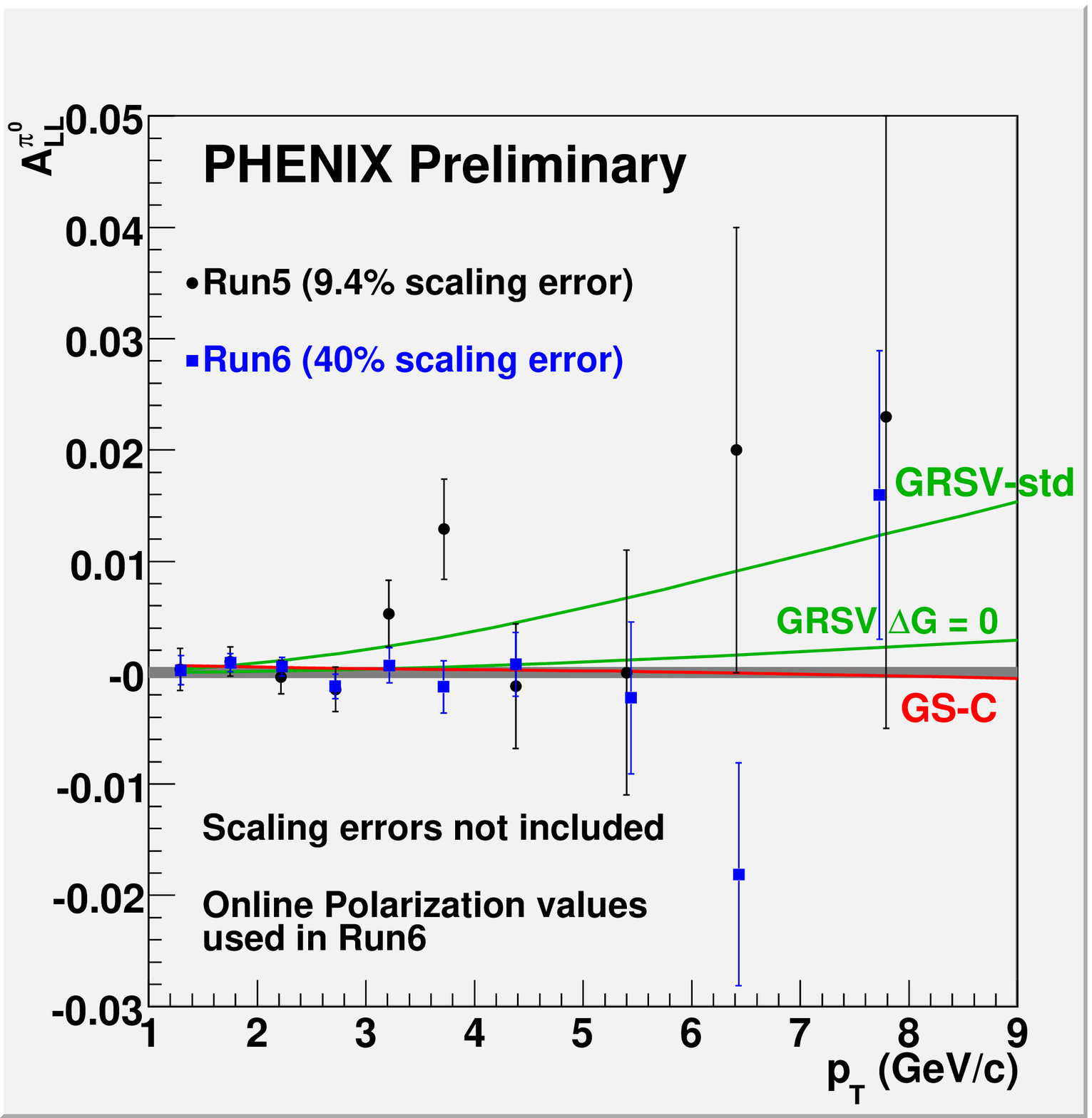}
\caption{Double helicity asymmetry for mid-rapidity inclusive jet (left
  panel) production from the 2006 (preliminary, see Ref.~\cite{star_jets_run5}
  for final results from 2005 data) data and $\pi^0$ (right
  panel) production from the 2006 
(preliminary\protect\footnotemark[1])
and 2005 data~\cite{allpi0} at $\sqrt{s}=200$~GeV as a
  function of $p_T$. The results for the asymmetries are compared to pQCD
  calculations \cite{marco} based on different sets of polarized
  PDFs~\cite{GRSV,GS}. See text for details.}
\label{asys}
\end{figure}
\begin{figure}[t]
\includegraphics[width=0.495\columnwidth]{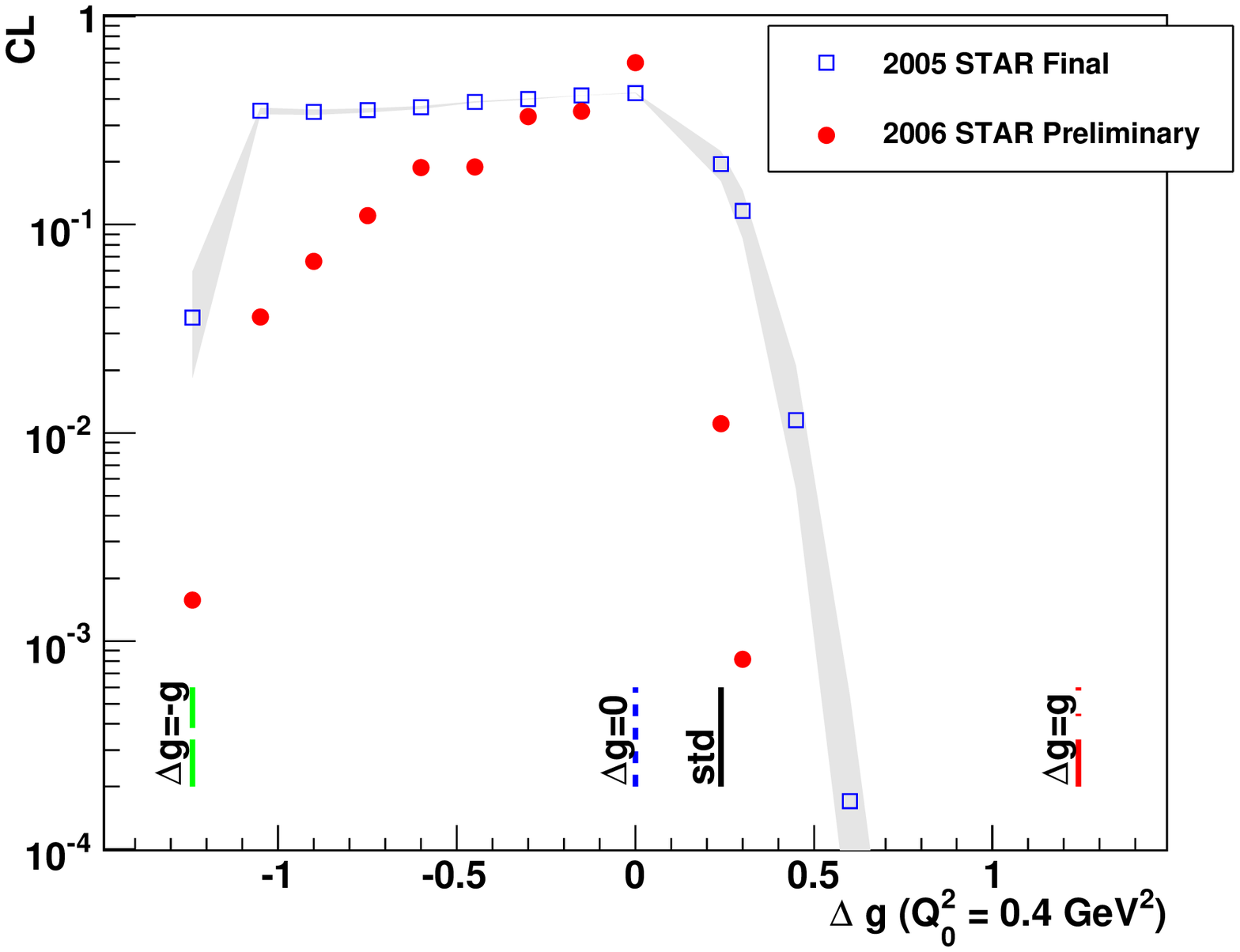}
\includegraphics[width=0.495\columnwidth,height=7.5cm]{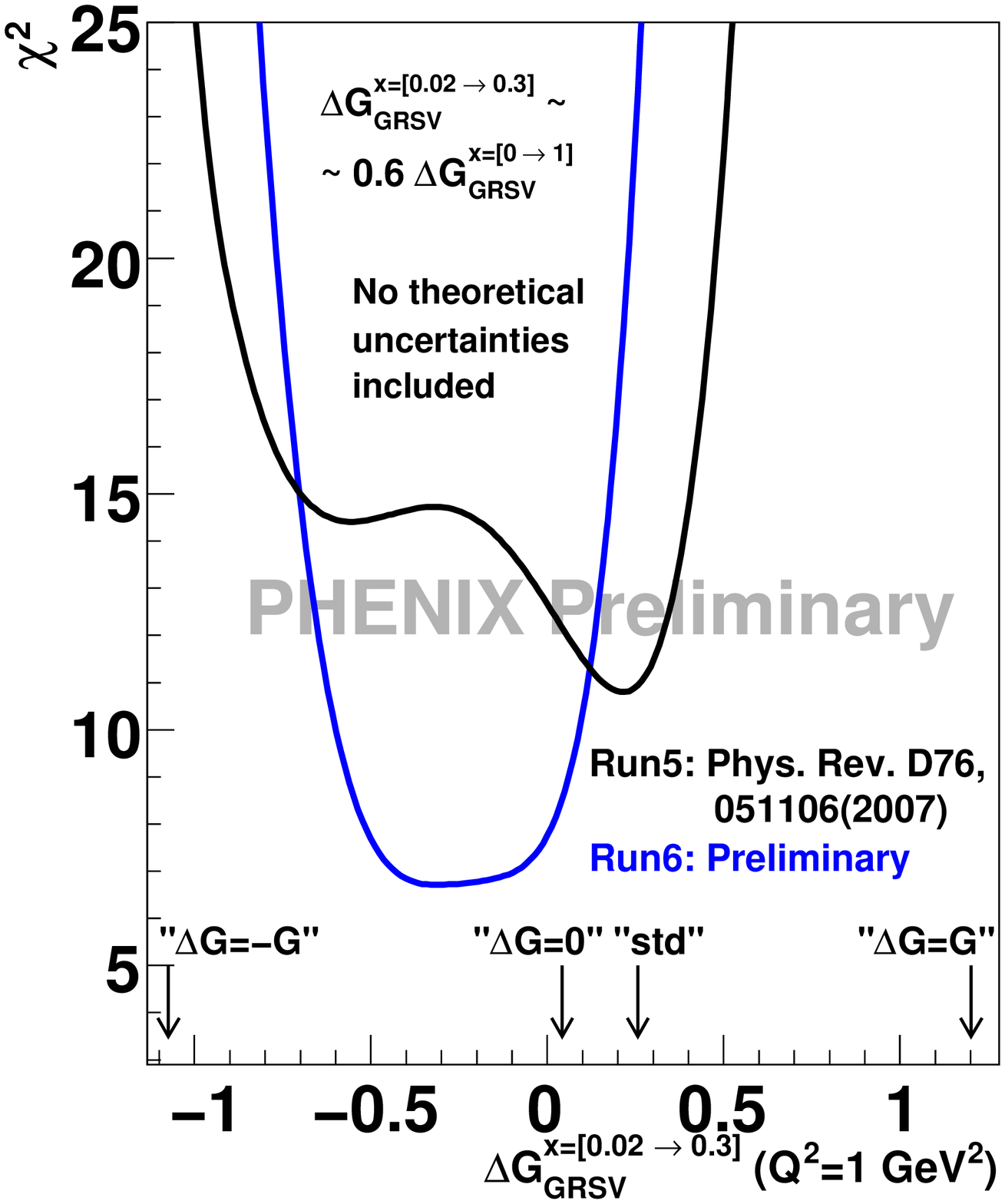}
\caption{Confidence level of the STAR jet data (left panel) and $\chi^2$ profile
for the PHENIX $\pi^0$ data (right panel) resulting from a comparison to 
four GRSV~\cite{GRSV} scenarios as well as for several other 
values of $\Delta G$ at the input scale~\cite{marco}.}
\label{chi2}
\end{figure}

\footnotetext[1]{Final results are now available in Ref.~\cite{run6_final}.}
The double helicity asymmetry as
a function of the transverse momentum $p_T$ for jet production at STAR and
$\pi^0$ production at PHENIX are shown in Fig.~\ref{asys}.
The results are compared to NLO pQCD calculations \cite{marco}, using
two sets of polarized PDFs.
In addition to the GS-C PDFs~\cite{GS}, the GRSV PDFs~\cite{GRSV} (GRSV-std) 
have been modified by assuming the polarized gluon PDF
to be zero ($\Delta G = 0$), equal ($\Delta G =G$), or opposite
($\Delta G = -G$) to the unpolarized gluon PDF at the input scale.
It is apparent from Fig.~\ref{asys} that the maximum and minimum GRSV scenarios, which are not even
shown anymore on the PHENIX plot, are ruled out by both experiments.
While not excluded, GRSV-std is disfavored by the data as can be
seen in Fig.~\ref{chi2},
where for the STAR jet data (PHENIX $\pi^0$ data) the confidence level ($\chi^2$
profile) is shown for the four GRSV scenarios as well as for several other 
values of $\Delta G$ at the input scale. 
Note that no uncertainties have been calculated for the GRSV
set of PDFs,
and that they in general are rather large for the poorly constrained $\Delta g(x)$ from
the DIS data as discussed above (see, e.g., Ref.~\cite{bb}), hence the RHIC results 
remain consistent with the results from DIS.

Based on Fig.~\ref{chi2} a small or maybe negative contribution of the gluon 
spin $\Delta G$ to the nucleon spin appears to be most
likely; however, this is a model dependent result.
The present inclusive measurements cannot determine $x$ but are sensitive to the 
gluon contribution integrated over the range $0.05 < x < 0.2$. Thus, in case
the gluon contribution is large outside the presently accessible $x$--range it could
still account for all the ``missing'' spin, even though the contribution in
the limited accessible $x$ range is rather small.
An example is the predicted asymmetry based on the GS-C PDF also shown in both 
panels of Fig.~\ref{asys}. It is
consistent with the present data, but the GS-C PDF gives a large contribution of gluons 
to the nucleon spin due to the fact that it is large and positive at small $x$ and becomes negative above $x
\approx 0.1.$ In other words, the gluon contribution is small in the measured region
since it has a node, but the positive contribution at smaller $x$ could account for
all the spin due to the abundance of gluons at small $x$.

\begin{figure}[t]
\includegraphics[width=0.495\columnwidth]{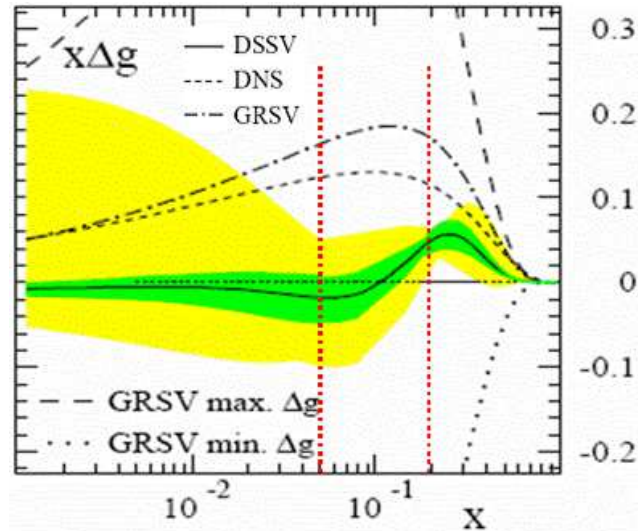}
\caption{Polarized gluon PDF from a fit~\cite{DSSV} that included for the first time
the pp data (in addition to DIS and semi-inclusive DIS), compared to the
earlier fits GRSV and DNS~\cite{deflorian}. Alternative DSSV fits corresponding to 
$\Delta \chi^2 = 1$ ($\Delta \chi^2 / \chi^2 =2\%$) are given by the 
green (yellow) bands. Figure taken from Ref.~\cite{DSSV}.}
\label{dssv}
\end{figure}

These features can also be seen in Fig.~\ref{dssv}, which shows a recent NLO
pQCD fit (DSSV~\cite{DSSV}), including for the first time
the pp data in addition to DIS and semi-inclusive DIS in the extraction of
the polarized PDFs,
a so-called global fit. 
In the $x$ range accessible at RHIC the DSSV fit yields smaller values
for $x \Delta g(x)$ than the earlier GRSV or DNS~\cite{deflorian} fits with much reduced
uncertainties. However, at smaller values of $x$ the uncertainties are large
and thus allow for a large $x \Delta g(x)$.
A lower (higher) range in
$x$ can be probed running RHIC at higher (lower) energies. Results
on $\pi^0$ production at PHENIX at $\sqrt{s}=62.4$~GeV
are already available~\cite{62}, while
running at $\sqrt{s}=500$~GeV is foreseen to start in 2009. In addition,
measurements at forward rapidity will allow access to a lower
range in $x$. 
Also, with more
data available in the future it is possible to advance from inclusive
measurements, in which only one particle in the final state is detected,
to measurements in which two or more leading particles or jets are detected in the final
state, allowing for a much better handle on the $x$-dependence.
Since in DIS the momentum fraction $x$ can be measured directly,
a precise measurement of the functional from of the polarized gluon PDF
$\Delta g(x)$ will be possible at a proposed polarized electron-proton
collider called EIC~\cite{eic}, which would be able to 
overcome the difficulties arising from the low energy scale and limited $Q^2$
range at fixed-target experiments.


This work is supported in part by the US Department of Energy.





\begin{thebibliography}{99}


\bibitem{emc} [European Muon Collaboration], J. Ashman et al., Phys. Lett. B
  206 (1988) 364.

\bibitem{hermes_g1} [HERMES Collaboration], A. Airapetian et al., Phys. Rev. D
  75 (2007) 012007. 

\bibitem{compass_g1_d} [COMPASS Collaboration], V.Yu. Alexakhin et al.,
  Phys. Lett. B 647 (2007) 8.

\bibitem{star_jets_run34} [STAR Collaboration], B.I. Abelev et al., 
Phys. Rev. Lett. 97 (2006) 252001.

\bibitem{allpi0} [PHENIX Collaboration], A. Adare et al., Phys. Rev. D 76
  (2007) 051106.


\bibitem{pC} O. Jinnouchi et al., RHIC/CAD Note 171 (2004).

\bibitem{jet} H. Okada et al., hep-ex/0601001.

\bibitem{star_jets_run5} [STAR Collaboration], B.I. Abelev et al., Phys. Rev. Lett. 100
  (2008) 232003.

\bibitem{run6_final} [PHENIX Collaboration], A. Adare et al., arXiv:0810.0694.

\bibitem{marco} M. Stratmann and W. Vogelsang, private communication.

\bibitem{GS} T. Gehrmann and W. J. Stirling, Phys. Rev. D 53 (1996) 6100.

\bibitem{GRSV} M. Gl\"uck, E. Reya, M. Stratmann, and W. Vogelsang,
  Phys. Rev. D 63 (2001) 094005.


\bibitem{bb} J. Bl\"umlein, H. B\"ottcher, Nucl. Phys. B 636 (2002) 225.


\bibitem{DSSV} D. de Florian, R. Sassot, M. Stratmann, and W. Vogelsang, 
Phys. Rev. Lett. 101 (2008) 072001.


\bibitem{deflorian} D. de Florian, G.A. Navarro, and R. Sassot, Phys. Rev. D
  71 (2005) 094018.

\bibitem{62} [PHENIX Collaboration], A. Adare et al., arXiv:0810.0701.


\bibitem{eic} A. Deshpande, R. Milner, R. Venugopalan, and W. Vogelsang,
  Ann. Rev. Nucl. Part. Sci. 55 (2005) 165; \verb$http://web.mit.edu/eicc/DOCUMENTS/EIC_LRP-20070424.pdf$



\end{thebibliography}
\end{document}